\documentclass[12pt]{article}
\usepackage{epsfig,cite,feynarts}

\input paperdef

\begin{document}
\thispagestyle{empty}

\def\thefootnote{\fnsymbol{footnote}}

\begin{flushright}
CERN-TH/2003-309\\
DCPT/03/148\\
DESY-03-153\\
IPPP/03/74\\
LMU 28/03\\
hep-ph/0312264
\end{flushright}

\begin{center}

{\large\sc {\bf Two-Loop SUSY Corrections to the}}

\vspace{0.4cm}

{\large\sc {\bf Anomalous Magnetic Moment of the Muon}}
 
\vspace{1cm}

{\sc 
S.~Heinemeyer$^{1,2}$%
\footnote{email: Sven.Heinemeyer@cern.ch}%
, D.~St\"ockinger$^{3,4}$%
\footnote{email: Dominik.Stockinger@durham.ac.uk}%
~and G.~Weiglein$^{4}$%
\footnote{email: Georg.Weiglein@durham.ac.uk}
}

\vspace*{1cm}

{\sl
$^1$ CERN, TH Division, 1211 Geneva 23, Switzerland

\vspace*{0.4cm}

$^2$Institut f\"ur theoretische Elementarteilchenphysik,
LMU M\"unchen, Theresienstr.\ 37, D--80333 M\"unchen, Germany

\vspace*{0.4cm}

$^3$ DESY Theorie, Notkestr. 85, 22603 Hamburg, Germany

\vspace*{0.4cm}

$^4$Institute for Particle Physics Phenomenology, University of Durham,\\
Durham DH1~3LE, UK
}

\end{center}

\vspace*{0.2cm}

\begin{abstract}

We calculate supersymmetric 
\twol\ corrections to the anomalous 
magnetic moment of the muon, consisting of diagrams with a closed
scalar fermion or fermion loop and gauge and/or Higgs boson exchange. We
discuss the numerical impact of each subclass of diagrams and
determine the leading contributions.
We analyze in detail constraints from experimental information on the
Higgs boson mass, 
$\De\rho$, and the branching ratios of $B \to X_s\ga$ and
$B_s\to\mu^+\mu^-$. 
If these constraints are taken into account, the largest possible
effect of our two-loop corrections is reduced from more than
$3\,\si$ (in terms of the current
experimental error) to $\sim0.5\,\si$, such that the influence 
on the total supersymmetric prediction is smaller than previously estimated.
However, exceptions arise in rather extreme parameter
scenarios with a strong non-universality between the soft 
breaking parameters in the stop and sbottom sectors.
\end{abstract}

\def\thefootnote{\arabic{footnote}}
\setcounter{page}{0}
\setcounter{footnote}{0}

\newpage


\section{Introduction}

A new era of precision measurements of the anomalous magnetic moment
of the muon $\amu \equiv (g_\mu - 2)/2$ has been initiated by 
the ``Muon g-2 Experiment'' (E821) at BNL, leading to the current
experimental world average of~\cite{g-2exp}
\BE
\amuexp = (11\, 659\, 208 \pm 6) \times 10^{-10}~.
\EE

The most recent
$e^+e^-$ data driven evaluations of the hadronic contributions by
\citeres{DEHZ,g-2HMNT,Jegerlehner} lead to the following Standard Model
(SM) predictions (the deviation from the experimental result is also
shown)%
\footnote{
The numbers for the combination of the experimental and the theory
error and the corresponding deviaton in terms of $\si$ have been
recalculated according to the new experimental result.
}%
:
\BEA
\amutheo &=& (11\, 659\, 180.9 \pm 8.0) \times 10^{-10} 
             \quad (27.1 \pm 10.0 ~:~2.7\,\si)
             \mbox{\cite{DEHZ} } 
\non \\
\amutheo &=& (11\, 659\, 175.6 \pm 7.5) \times 10^{-10} 
             \quad (32.4 \pm ~\,9.6 ~:~3.3\,\si)
             \mbox{\cite{g-2HMNT} }
 \non \\
\amutheo &=& (11\, 659\, 179.4 \pm 9.3) \times 10^{-10} 
             \quad (28.6 \pm 11.1 ~:~2.5\,\si)
             \mbox{\cite{Jegerlehner} } . \non
\EEA
Recent analyses concerning $\tau$ data indicate that uncertainties due to
isospin breaking effects may have been underestimated
earlier~\cite{Jegerlehner}, so that 
with a better theoretical understanding of the isospin breaking effects
the $\tau$-based results could come closer to the $e^+e^-$- based
results. One may thus hope that eventually a combination of $e^+e^-$ and $\tau$
data will lead to an even more
precise theoretical prediction. 

While the present $\sim 2.5 - 3.3\,\si$ deviation between 
the SM prediction for $\amu$
and the experimental result can of course not be regarded as strong evidence
for new physics, an increased accuracy of both theory and experiment
might give rise to a significantly larger deviation in the future.
On the other hand, already the current precision leads to very
restrictive bounds on new physics scenarios.

This is illustrated by the fact that the experimental precision of
$6\times 10^{-10}$ has now reached the level of
the standard electroweak and of typical supersymmetric (SUSY)
contributions. The 
electroweak one- and two-loop (and higher-order)
contributions in the SM amount to 
$19.5 \times 10^{-10}$ and $-4.1 \times 10^{-10}$, respectively, see
\citeres{g-2review,g-2review2} for reviews.
The SUSY contributions
are generally suppressed by $\MW^2/\tilde M^2$, where $\MW$ is the
mass of the $W$~boson and $\tilde M$ is
the typical scale of the SUSY particle masses. However, for large
values of $\tb$, the ratio of the vacuum expectation values of the two
Higgs doublets in the Minimal Supersymmetric Standard Model (MSSM),
the muon Yukawa coupling is enhanced by $\tb$ as compared to the SM.
The supersymmetric one-loop contribution is approximately given by
\cite{g-2MSSMf1l}
\BE
|\amu^{\SU}| = 13 \times 10^{-10} 
             \KL \frac{100 \gev}{\tilde M} \KR^2 \tb,
\EE
where all SUSY masses are assumed to be equal to $\tilde M$.
The involved SUSY particles are neutralinos, charginos and scalar
leptons of the second generation.
The magnitude of the supersymmetric \onel\ contribution is at the right level 
to account for the $\sim 3\si$ deviation between the SM prediction and the 
data, and even larger shifts are possible.
The supersymmetric two-loop contributions are known only in some
approximations. Since the \onel\ contribution can be large, the
\twol\ corrections can be expected to be quite
important, even beyond the leading QED-logarithms
\cite{g-2MSSMlog2l}.

The importance of the SUSY \twol\ contributions is 
twofold. On the one hand, their inclusion increases the accuracy of
the bounds on the supersymmetric parameter space (see e.g.\
\citeres{g-2appl1,g-2appl2}). On the other hand, 
the supersymmetric \twol\
contributions depend on many additional parameters and can in principle
be large even if the \onel\ diagrams are suppressed due to heavy smuons and
sneutrinos. 

Particularly interesting contributions are the ones enhanced by large
values of the Higgs mixing parameter $\mu$ and a large trilinear
coupling $A$ (where $A$ generically denotes the Higgs--stop or
Higgs--sbottom coupling, $A_{t,b}$). They arise from so-called 
Barr-Zee \twol\ diagrams where a Higgs boson is exchanged between the
external muon and a 3rd generation sfermion loop. 
Results for such contributions were
obtained in \citeres{g-2BarrZee1,g-2BarrZee2} and found to give huge
contributions up to \order{20\times10^{-10}} if $\tb$ is large
and $\mu,A$ are of the order of several TeV. In these analyses, however,
other experimental constraints on the parameter space of the MSSM were
neglected.
Moreover, the results of \citeres{g-2BarrZee1,g-2BarrZee2} for the $H^\pm$
contribution disagree by a factor 4, so that an independent check seems
to be necessary.

In this paper we present a calculation and numerical analysis of all
\twol\ contributions $\delamu$ in the MSSM where a closed 3rd generation
sfermion or fermion loop is inserted into a \onel\ diagram with
gauge-boson and/or Higgs-boson exchange. 
This set of diagrams contains the terms 
$\propto \mu,A$ but also 
other terms enhanced by the large Yukawa couplings of the $t$, and
(for large $\tb$) $b$, $\tau$, as well as terms without any
enhancement. All of these contributions are included in our final
result.

Our numerical analysis is focused on two questions: 
what are the numerical results for the individual subclasses, and which of
them should be taken into account for a reliable supersymmetric
prediction for $\amu$? 
Secondly, are huge two-loop contributions of
\order{20\times10^{-10}} still possible if existing experimental
constraints on the
supersymmetric parameter space are taken into account?

The rest of the paper is organized as follows. 
In \refse{sec:calc} we present the \twol\ diagrams and the method of their
evaluation. The numerical analysis
is given in \refses{sec:numeval},~\ref{sec:expconst},~\ref{sec:nonuniv}. 
The importance of this class of \twol\ corrections and
their numerical size is discussed in \refse{sec:numeval}, taking into account
constraints on the SUSY parameter space from other experimental
information. The leading contributions and
the influence of the individual experimental constraints are examined
in \refse{sec:expconst}. Effects from non-universality of
the soft SUSY-breaking parameters are analyzed in
\refse{sec:nonuniv}.  We conclude with \refse{sec:conclusions}.


\section{Calculation}
\label{sec:calc}

In this section we briefly describe the diagrams we have investigated, their
evaluation and the tools that have been used. 

The set of diagrams calculated in this paper corresponds to the 
fermion/sfermion corrections to non-supersymmetric, i.e.\ two Higgs doublet 
model type, contributions. It forms a gauge-independent class of
diagrams. In order to discuss the shift between the
MSSM and the SM predictions, we subtract the pure SM contribution from
our result (where the SM Higgs boson mass $\MHSM$ is set to the
value of the lightest MSSM Higgs boson mass, $\Mh$).

The one-loop diagrams corresponding to the contributions considered in
this paper are the SM one-loop diagrams.
Expressing the one-loop result in terms of the Fermi constant
$G_{\mu}$ and $\sw^2 \equiv 1 - \MW^2/\MZ^2$, ($\MZ$ being the
$Z$~boson mass), it takes the conventional form of the
electroweak one-loop result in the SM (omitting the QED
contribution) \cite{g-2review,g-2review2},
\BE
a_{\mu}^{\rm EW,1L} = \frac{G_{\mu}}{8 \pi^2 \sqrt{2}}
m_{\mu}^2 \left[\frac{5}{3} + \frac{1}{3} (1 - 4 \sw^2)^2\right] .
\label{eq:sm1loop}
\EE

\begin{figure}[htb!]
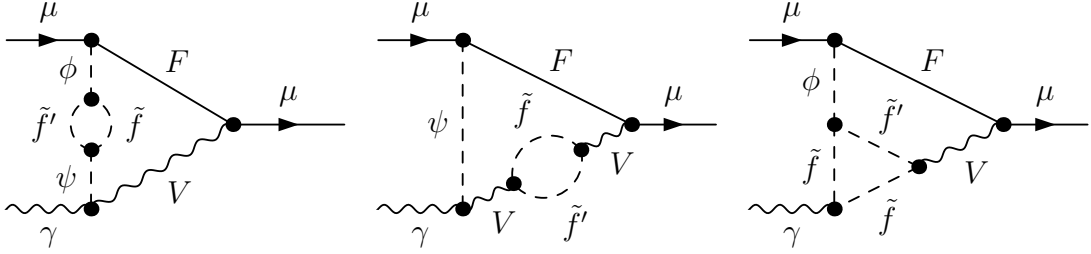

\begin{center}
\unitlength=1.bp%
\begin{feynartspicture}(432,140)(3,1)

\FADiagram{}
\FAProp(0.,15.)(5.,15.)(0.,){/Straight}{1}
\FALabel(2.5,16.07)[b]{$\mu$}
\FAProp(0.,5.)(5.,5.)(0.,){/Sine}{0}
\FALabel(2.5,3.93)[t]{$\gamma$}
\FAProp(20.,10.)(13.4,10.)(0.,){/Straight}{-1}
\FALabel(16.7,11.07)[b]{$\mu$}
\FAProp(5.,15.)(13.4,10.)(0.,){/Straight}{0}
\FALabel(9.38493,13.1371)[bl]{$F$}
\FAProp(5.,11.5)(5.,15.)(0.,){/ScalarDash}{0}
\FALabel(4.18,13.25)[r]{$\phi$}
\FAProp(5.,8.5)(5.,11.5)(0.8,){/ScalarDash}{0}
\FALabel(7.02,10.)[l]{$\Sferm$}
\FAProp(5.,8.5)(5.,11.5)(-0.8,){/ScalarDash}{0}
\FALabel(2.98,10.)[r]{$\Sfermp$}
\FAProp(5.,8.5)(5.,5.)(0.,){/ScalarDash}{0}
\FALabel(4.18,6.75)[r]{$\psi$}
\FAProp(5.,5.)(13.4,10.)(0.,){/Sine}{0}
\FALabel(9.5128,6.6481)[tl]{$V$}
\FAVert(5.,8.5){0}
\FAVert(5.,11.5){0}
\FAVert(5.,15.){0}
\FAVert(5.,5.){0}
\FAVert(13.4,10.){0}

\FADiagram{}
\FAProp(0.,15.)(5.,15.)(0.,){/Straight}{1}
\FALabel(2.5,16.07)[b]{$\mu$}
\FAProp(0.,5.)(5.,5.)(0.,){/Sine}{0}
\FALabel(2.5,3.93)[t]{$\gamma$}
\FAProp(20.,10.)(15.,10.)(0.,){/Straight}{-1}
\FALabel(17.5,11.07)[b]{$\mu$}
\FAProp(5.,15.)(5.,5.)(0.,){/ScalarDash}{0}
\FALabel(4.18,10.)[r]{$\psi$}
\FAProp(5.,15.)(15.,10.)(0.,){/Straight}{0}
\FALabel(10.1014,13.1828)[bl]{$F$}
\FAProp(8.,6.5)(5.,5.)(0.,){/Sine}{0}
\FALabel(6.71318,4.84364)[tl]{$V$}
\FAProp(12.,8.5)(8.,6.5)(0.8,){/ScalarDash}{0}
\FALabel(9.09862,9.78276)[br]{$\Sferm$}
\FAProp(12.,8.5)(8.,6.5)(-0.8,){/ScalarDash}{0}
\FALabel(10.9014,5.21724)[tl]{$\Sfermp$}
\FAProp(12.,8.5)(15.,10.)(0.,){/Sine}{0}
\FALabel(13.7132,8.34364)[tl]{$V$}
\FAVert(12.,8.5){0}
\FAVert(8.,6.5){0}
\FAVert(5.,15.){0}
\FAVert(5.,5.){0}
\FAVert(15.,10.){0}

\FADiagram{}
\FAProp(0.,15.)(5.,15.)(0.,){/Straight}{1}
\FALabel(2.5,16.07)[b]{$\mu$}
\FAProp(0.,5.)(5.,5.)(0.,){/Sine}{0}
\FALabel(2.5,3.93)[t]{$\gamma$}
\FAProp(20.,10.)(15.,10.)(0.,){/Straight}{-1}
\FALabel(17.5,11.07)[b]{$\mu$}
\FAProp(5.,15.)(15.,10.)(0.,){/Straight}{0}
\FALabel(10.1014,13.1828)[bl]{$F$}
\FAProp(5.,10.)(5.,15.)(0.,){/ScalarDash}{0}
\FALabel(4.18,12.5)[r]{$\phi$}
\FAProp(5.,10.)(5.,5.)(0.,){/ScalarDash}{0}
\FALabel(4.18,7.5)[r]{$\Sferm$}
\FAProp(10.,7.5)(5.,10.)(0.,){/ScalarDash}{0}
\FALabel(7.60138,9.43276)[bl]{$\Sfermp$}
\FAProp(10.,7.5)(5.,5.)(0.,){/ScalarDash}{0}
\FALabel(7.60138,5.56724)[tl]{$\Sferm$}
\FAProp(10.,7.5)(15.,10.)(0.,){/Sine}{0}
\FALabel(12.7132,7.84364)[tl]{$V$}
\FAVert(10.,7.5){0}
\FAVert(5.,10.){0}
\FAVert(5.,15.){0}
\FAVert(5.,5.){0}
\FAVert(15.,10.){0}

\end{feynartspicture}

\caption{%
Some generic \twol\ SUSY diagrams of type ($\sfn V \phi$) involving (at least)
one gauge and one Higgs boson and a closed scalar fermion loop. 
$F = \mu, \bar\nu_\mu$; 
$\phi = h, H, A, H^\pm, G, G^\pm$; 
$\psi = G^\pm$; 
$\Sferm, \Sfermp = \Stop, \Sbot, \Stau, \Snet$;
$V = \ga, Z, W$. Further diagrams of this type are obtained by
contracting the $\psi$ line in the first diagram or by interchanging
Higgs and vector bosons.
}
\label{fig:HGsf}
\end{center}
\end{figure}

\begin{figure}[ht!]
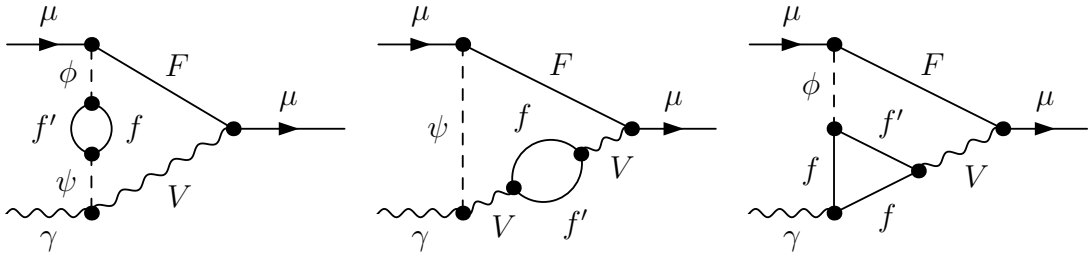

\begin{center}
\unitlength=1.bp%
\begin{feynartspicture}(432,140)(3,1)

\FADiagram{}
\FAProp(0.,15.)(5.,15.)(0.,){/Straight}{1}
\FALabel(2.5,16.07)[b]{$\mu$}
\FAProp(0.,5.)(5.,5.)(0.,){/Sine}{0}
\FALabel(2.5,3.93)[t]{$\gamma$}
\FAProp(20.,10.)(13.4,10.)(0.,){/Straight}{-1}
\FALabel(16.7,11.07)[b]{$\mu$}
\FAProp(5.,15.)(13.4,10.)(0.,){/Straight}{0}
\FALabel(9.38493,13.1371)[bl]{$F$}
\FAProp(5.,11.5)(5.,15.)(0.,){/ScalarDash}{0}
\FALabel(4.18,13.25)[r]{$\phi$}
\FAProp(5.,8.5)(5.,11.5)(0.8,){/Straight}{0}
\FALabel(7.02,10.)[l]{$f$}
\FAProp(5.,8.5)(5.,11.5)(-0.8,){/Straight}{0}
\FALabel(2.98,10.)[r]{$f'$}
\FAProp(5.,8.5)(5.,5.)(0.,){/ScalarDash}{0}
\FALabel(4.18,6.75)[r]{$\psi$}
\FAProp(5.,5.)(13.4,10.)(0.,){/Sine}{0}
\FALabel(9.5128,6.6481)[tl]{$V$}
\FAVert(5.,8.5){0}
\FAVert(5.,11.5){0}
\FAVert(5.,15.){0}
\FAVert(5.,5.){0}
\FAVert(13.4,10.){0}

\FADiagram{}
\FAProp(0.,15.)(5.,15.)(0.,){/Straight}{1}
\FALabel(2.5,16.07)[b]{$\mu$}
\FAProp(0.,5.)(5.,5.)(0.,){/Sine}{0}
\FALabel(2.5,3.93)[t]{$\gamma$}
\FAProp(20.,10.)(15.,10.)(0.,){/Straight}{-1}
\FALabel(17.5,11.07)[b]{$\mu$}
\FAProp(5.,15.)(5.,5.)(0.,){/ScalarDash}{0}
\FALabel(4.18,10.)[r]{$\psi$}
\FAProp(5.,15.)(15.,10.)(0.,){/Straight}{0}
\FALabel(10.1014,13.1828)[bl]{$F$}
\FAProp(8.,6.5)(5.,5.)(0.,){/Sine}{0}
\FALabel(6.71318,4.84364)[tl]{$V$}
\FAProp(12.,8.5)(8.,6.5)(0.8,){/Straight}{0}
\FALabel(9.09862,9.78276)[br]{$f$}
\FAProp(12.,8.5)(8.,6.5)(-0.8,){/Straight}{0}
\FALabel(10.9014,5.21724)[tl]{$f'$}
\FAProp(12.,8.5)(15.,10.)(0.,){/Sine}{0}
\FALabel(13.7132,8.34364)[tl]{$V$}
\FAVert(12.,8.5){0}
\FAVert(8.,6.5){0}
\FAVert(5.,15.){0}
\FAVert(5.,5.){0}
\FAVert(15.,10.){0}

\FADiagram{}
\FAProp(0.,15.)(5.,15.)(0.,){/Straight}{1}
\FALabel(2.5,16.07)[b]{$\mu$}
\FAProp(0.,5.)(5.,5.)(0.,){/Sine}{0}
\FALabel(2.5,3.93)[t]{$\gamma$}
\FAProp(20.,10.)(15.,10.)(0.,){/Straight}{-1}
\FALabel(17.5,11.07)[b]{$\mu$}
\FAProp(5.,15.)(15.,10.)(0.,){/Straight}{0}
\FALabel(10.1014,13.1828)[bl]{$F$}
\FAProp(5.,10.)(5.,15.)(0.,){/ScalarDash}{0}
\FALabel(4.18,12.5)[r]{$\phi$}
\FAProp(5.,10.)(5.,5.)(0.,){/Straight}{0}
\FALabel(4.18,7.5)[r]{$f$}
\FAProp(10.,7.5)(5.,10.)(0.,){/Straight}{0}
\FALabel(7.60138,9.43276)[bl]{$f'$}
\FAProp(10.,7.5)(5.,5.)(0.,){/Straight}{0}
\FALabel(7.60138,5.56724)[tl]{$f$}
\FAProp(10.,7.5)(15.,10.)(0.,){/Sine}{0}
\FALabel(12.7132,7.84364)[tl]{$V$}
\FAVert(10.,7.5){0}
\FAVert(5.,10.){0}
\FAVert(5.,15.){0}
\FAVert(5.,5.){0}
\FAVert(15.,10.){0}

\end{feynartspicture}

\caption{%
Generic \twol\ SUSY diagrams of type ($f V \phi$) involving (at least)
one gauge and one Higgs boson and a closed SM fermion loop. 
$F = \mu, \bar\nu_\mu$; 
$\phi = h, H, A, H^\pm, G, G^\pm$; 
$\psi = G^\pm$; 
$f, f' = t, b, \tau, \nu_\tau$; 
$V = \ga, Z, W$.
}
\label{fig:HGf}
\end{center}
\vspace{-1.5em}
\end{figure}

The two-loop diagrams that we calculate can be subdivided into three classes:
\\{($\sfn V \phi$)} 
diagrams with a sfermion ($\Stop$, $\Sbot$, $\Stau$,
$\Snet$) loop, where at least one gauge and one Higgs boson are
exchanged, see \reffi{fig:HGsf};
\\{($\sfn V V$)} diagrams with a
sfermion loop, where only gauge bosons appear in the second loop,
see \reffi{fig:Gsf};
\\{($f V \phi$)} diagrams with a fermion ($t$, $b$, $\tau$, $\nu_\tau$)
loop, where at least one gauge and one Higgs boson are
present in the other loop, see \reffi{fig:HGf}. The corresponding
diagrams with only gauge bosons are identical to the SM diagrams and
give no genuine SUSY contribution.

For our later analysis we further split up the ($\sfn V \phi$)
diagrams of Fig.\ \ref{fig:HGsf} into the following groups:
diagrams with photon and Higgs exchange ($\sfn \ga \{h,H\}$),
$Z$/Higgs exchange ($\sfn Z \{h,H\}$), and $W$/charged Higgs exchange
($\sfn W^\pm H^\mp$). The remaining sfermion loop diagrams containing
only gauge boson and Goldstone boson exchange  are grouped together
with the diagrams involving only gauge 
bosons, ($\sfn W^\pm G^\mp$)+($\sfn VV$). All these groups are
separately gauge independent in $R_\xi$-gauges at the order $m_\mu^2/M_W^2$.
Note that since we neglect $\cp$-violating phases,  diagrams with
photon or $Z$ and $\cp$-odd Higgs bosons $A^0$, $G^0$ do not
contribute. The diagrams ($\sfn \ga \{h,H\}$) and ($\sfn W^\pm H^\mp$)
are the ones evaluated in \citeres{g-2BarrZee1,g-2BarrZee2} neglecting
all but the leading terms in the sfermion--Higgs couplings.

All diagrams are understood to include the corresponding subloop
renormalization. For the fermion loop class ($fV\phi$) we actually
calculate the difference between the 
Standard Model and the MSSM, which originates from the extended Higgs
sector of the MSSM. Diagrams where two Higgs bosons couple to
the external muon are suppressed by an extra factor of $m_\mu^2/\MW^2$ and
hence negligible.

In order to perform a systematic calculation, 
a {\em Mathematica} program has been written
that can deal with all kinds of MSSM \twol\ contributions to
$\amu$. Its main steps are the following: 
The amplitudes for
$\amu$ are generated using the program
\fa~\cite{feynarts,fa-mssm}, and the appropriate
projector~\cite{g-2SM2lA,g-2SM2lB} is applied. The Dirac algebra and the
conversion to a linear combination of \twol\ integrals is performed
using \tc~\cite{2lred}. 
In order to simplify the integrals, a large mass
expansion \cite{smirnov} is applied where the muon mass is
taken as small and all other masses as large. All resulting \twol\
integrals are either \twol\ vacuum integrals or products of \onel\
integrals. They can be 
reduced to the standard integrals $T_{134}$~\cite{t134} and $A_0$ and
$B_0$~\cite{a0b0c0d0} and can be evaluated analytically.
The asymptotic expansion has to be performed up to terms of
order $m_\mu^2$. Terms of lower power in $m_\mu/M_{\rm heavy}$ (where
$M_{\rm heavy}$ represents all kinds of other masses) cancel
each other as required for non-QED corrections. Terms of higher powers
in $m_\mu/M_{\rm heavy}$ are numerically irrelevant and can be safely
neglected.

\begin{figure}[ht!]
\begin{center}
\unitlength=1.bp%
\begin{feynartspicture}(432,140)(3,1)

\FADiagram{}
\FAProp(0.,15.)(5.,15.)(0.,){/Straight}{1}
\FALabel(2.5,16.07)[b]{$\mu$}
\FAProp(0.,5.)(5.,5.)(0.,){/Sine}{0}
\FALabel(2.5,3.93)[t]{$\gamma$}
\FAProp(20.,10.)(13.4,10.)(0.,){/Straight}{-1}
\FALabel(16.7,11.07)[b]{$\mu$}
\FAProp(5.,15.)(13.4,10.)(0.,){/Straight}{0}
\FALabel(9.38493,13.1371)[bl]{$F$}
\FAProp(5.,11.5)(5.,15.)(0.,){/Sine}{0}
\FALabel(3.93,13.25)[r]{$V$}
\FAProp(5.,8.5)(5.,11.5)(0.8,){/ScalarDash}{0}
\FALabel(7.02,10.)[l]{$\Sferm$}
\FAProp(5.,8.5)(5.,11.5)(-0.8,){/ScalarDash}{0}
\FALabel(2.98,10.)[r]{$\Sfermp$}
\FAProp(5.,8.5)(5.,5.)(0.,){/Sine}{0}
\FALabel(3.93,6.75)[r]{$V$}
\FAProp(5.,5.)(13.4,10.)(0.,){/Sine}{0}
\FALabel(9.5128,6.6481)[tl]{$V$}
\FAVert(5.,8.5){0}
\FAVert(5.,11.5){0}
\FAVert(5.,15.){0}
\FAVert(5.,5.){0}
\FAVert(13.4,10.){0}

\FADiagram{}
\FAProp(0.,15.)(5.,15.)(0.,){/Straight}{1}
\FALabel(2.5,16.07)[b]{$\mu$}
\FAProp(0.,5.)(5.,5.)(0.,){/Sine}{0}
\FALabel(2.5,3.93)[t]{$\gamma$}
\FAProp(20.,10.)(15.,10.)(0.,){/Straight}{-1}
\FALabel(17.5,11.07)[b]{$\mu$}
\FAProp(5.,15.)(5.,5.)(0.,){/Sine}{0}
\FALabel(3.93,10.)[r]{$V$}
\FAProp(5.,15.)(15.,10.)(0.,){/Straight}{0}
\FALabel(10.1014,13.1828)[bl]{$F$}
\FAProp(8.,6.5)(5.,5.)(0.,){/Sine}{0}
\FALabel(6.71318,4.84364)[tl]{$V$}
\FAProp(12.,8.5)(8.,6.5)(0.8,){/ScalarDash}{0}
\FALabel(9.09862,9.78276)[br]{$\Sferm$}
\FAProp(12.,8.5)(8.,6.5)(-0.8,){/ScalarDash}{0}
\FALabel(10.9014,5.21724)[tl]{$\Sfermp$}
\FAProp(12.,8.5)(15.,10.)(0.,){/Sine}{0}
\FALabel(13.7132,8.34364)[tl]{$V$}
\FAVert(12.,8.5){0}
\FAVert(8.,6.5){0}
\FAVert(5.,15.){0}
\FAVert(5.,5.){0}
\FAVert(15.,10.){0}

\FADiagram{}
\FAProp(0.,15.)(5.,15.)(0.,){/Straight}{1}
\FALabel(2.5,16.07)[b]{$\mu$}
\FAProp(0.,5.)(5.,5.)(0.,){/Sine}{0}
\FALabel(2.5,3.93)[t]{$\gamma$}
\FAProp(20.,10.)(15.,10.)(0.,){/Straight}{-1}
\FALabel(17.5,11.07)[b]{$\mu$}
\FAProp(5.,15.)(15.,10.)(0.,){/Straight}{0}
\FALabel(10.1014,13.1828)[bl]{$F$}
\FAProp(5.,10.)(5.,15.)(0.,){/Sine}{0}
\FALabel(3.93,12.5)[r]{$V$}
\FAProp(5.,10.)(5.,5.)(0.,){/ScalarDash}{0}
\FALabel(4.18,7.5)[r]{$\Sferm$}
\FAProp(10.,7.5)(5.,10.)(0.,){/ScalarDash}{0}
\FALabel(7.60138,9.43276)[bl]{$\Sfermp$}
\FAProp(10.,7.5)(5.,5.)(0.,){/ScalarDash}{0}
\FALabel(7.60138,5.56724)[tl]{$\Sferm$}
\FAProp(10.,7.5)(15.,10.)(0.,){/Sine}{0}
\FALabel(12.7132,7.84364)[tl]{$V$}
\FAVert(10.,7.5){0}
\FAVert(5.,10.){0}
\FAVert(5.,15.){0}
\FAVert(5.,5.){0}
\FAVert(15.,10.){0}

\end{feynartspicture}

\caption{%
Some generic \twol\ SUSY diagrams of type ($\sfn V V$) involving gauge
bosons and a closed scalar fermion loop.
$F = \mu, \bar\nu_\mu$; 
$\Sferm, \Sfermp = \Stop, \Sbot, \Stau, \Snet$;
$V = \ga, Z, W$.
Further diagrams of this typ involving four-point vertices exist as well.
}
\label{fig:Gsf}
\end{center}
\vspace{-1.5em}
\end{figure}

The counterterm diagrams contain the renormalization
constants $\de M^2_{W,Z}$, $\de Z_e$, $\de t_{h,H}$ corresponding
to mass, charge and tadpole renormalization and can be easily
evaluated. We choose the on-shell renormalization scheme
\cite{onshell}. This leads to $\de M^2_{W,Z} = {\rm
  Re}\Si^T_{W,Z}(M^2_{W,Z})$, where 
$\Si_{W,Z}^T$ 
denote the transverse parts of the gauge-boson self-energies. 
The charge renormalization is given by $\de Z_e = - 1/2\; \Sip_\ga(0)$,
where $\Sip$ denotes the derivative of the self-energy with respect to
the momentum squared.
The tadpoles are renormalized such that the sum of the tadpole
contribution $T$ and the counterterm vanishes, i.e.\
$\de t_{h,H} = -T_{h,H}$.

As mentioned above, see \refeq{eq:sm1loop}, we are using a \onel\ result
which is parametrized in terms of $G_{\mu}$ instead
of the ratio $\al/\MW^2$. Therefore our two-loop correction 
contains a term given by the product of the corresponding \onel\
result and $\De r$, where the latter denote the one-loop corrections
to muon decay, $\mu \to \nu_\mu \, e \, \bar\nu_e$.
Relevant here are only the contributions arising from 3rd
family sfermion loops to $\De r$. These corrections are included in
our \twol\ result (in the ($\sfn V V$) class).

As a cross check we have evaluated the SM \twol\ diagrams with a
closed fermion loop as presented in \citeres{g-2SM2lA,g-2SM2lB} 
and found perfect agreement separately for each diagram
(after going to the limit $\sw^2 \to 1/4$, used in
\citeres{g-2SM2lA,g-2SM2lB}). 
We have furthermore checked the UV-finiteness of our result as well as
the cancellation of the wave function renormalization constants.
We also found agreement with \citere{g-2BarrZee2} for the 
contributions to the ($\sfn \ga \{h,H\}$) and ($\sfn W^\pm H^\mp$)
diagrams calculated there. This confirms that the earlier result of
\citere{g-2BarrZee1} is too large by a factor of 4.

Our final result for the sum of all diagrams is rather
lengthy and not displayed here. It 
is included as a Fortran subroutine in the code \fh~\cite{feynhiggs} 
(see: {\tt www.feynhiggs.de}). It can also be obtained as a
{\em Mathematica} formula from the authors upon request.


\section{Numerical results allowed
  by experimental\\constraints for the different sets of diagrams } 
\label{sec:numeval}

The MSSM \twol\ contributions to $\amu$ depend on many parameters,
most notably on $\tan\beta$, the $\mu$ parameter, the trilinear soft
SUSY-breaking parameters $A_{t,b,\tau}$, the mass of the $\cp$-odd
Higgs $M_A$, and the 
soft SUSY-breaking parameters $M_{Q,L,U,D,E}$ appearing in the
sfermion mass
matrices. In \citeres{g-2BarrZee1,g-2BarrZee2} it was shown that in
particular large $\mu$ and $A$ parameters can give rise to very large
contributions of the ($\tilde{f}\gamma \{H,h\}$) and ($\tilde{f}W^\pm H^\mp$)
diagrams, however ignoring existing experimental constraints on the
MSSM parameter space. In order to find out the largest possible
contributions of each class of diagrams, we perform a scan of the MSSM
parameter space. We vary the parameters in the ranges
\BEA
-3 \tev ~\le &\mu& \le ~3 \tev \non \\
-3 \tev ~\le &A_{t,b}& \le ~3 \tev \non \\
150 \gev ~\le &\MA& \le ~1 \tev \non \\
0 ~\le &\msusy& \le ~1 \tev
\label{bounds}
\EEA
where we have set $\msusy = M_Q = M_L = M_U = M_D = M_E$ 
($M_Q, M_U$ are the soft SUSY-breaking parameters in the $\Stop$~mass
matrix, $M_Q, M_D$ in the $\Sbot$~mass matrix, and $M_L, M_E$ in the
$\Stau$~mass matrix)
and $A_\tau = \Ab$.
Furthermore we fix $\tb$ to $\tb=50$. Large values
of $\tb$ and small values of $\MA$ generically lead to larger SUSY
contributions but also to more restrictive experimental
constraints. These two effects tend to cancel each other. We have
checked that the maximum contributions from our diagrams to $\amu$ 
allowed by the experimental
constraints are about the same for $\tb = 25$, $\tb = 37$ and $\tb =
50$ and when $\MA$ is varied in the range $\MA = 90 \ldots 150
\gev$. The effect of relaxing the restriction of a common soft
SUSY-breaking parameter in the sfermion mass matrices will be
described in \refse{sec:nonuniv}. As SM input 
parameters we use $\mt = 175 \gev$, and $\mb(\mt) = 3 \gev$ (in order to absorb
leading QCD corrections).

\begin{table}[tb]
\begin{center}
\begin{tabular}{c|c|c|c|c}
Quantity & $\Mh$  & $ \De\rho^{\SU}$  & $ \br(B_s\to\mu^+\mu^-)$  &  
$\De_{B \to X_s\ga}$ \\
\hline
strong bound  &  $>111.4 \gev$  &  $<3\times10^{-10}$  
& $ < 0.97\times10^{-6}$ & $< 1.0\times10^{-4}$ \\
weak bound  &  $>106.4 \gev$  &  $<4\times10^{-10}$  
& $ < 1.2\times10^{-6}$ & $< 1.5\times10^{-4}$ 
\end{tabular}
\end{center}
\caption{Strong and weak bounds imposed on the MSSM parameter
  space. $\De_{B \to X_s\ga}=|\br(B \to X_s\ga)-3.34\times10^{-4}|$,
  where $3.34\times10^{-4}$ is the current experimental central value
  \cite{bsg}.}
\label{tab:bounds}
\end{table} 

We restrict the parameter space further by imposing the following
experimental constraints:%
\footnote{%
There are of course also lower bounds on sfermion masses from direct
searches. The bounds from LEP are roughly
$m_{\Stop,\Sbot}\gsim100\gev$. From Run I of the Tevatron stronger
bounds arise for parts of the MSSM parameter space
\cite{pdg,mschmitt}. We do not impose the direct
bounds explicitly in the scans since we present the results of $\amu$
as functions of the lightest sfermion mass.
}%

\begin{itemize}

\item The lightest $\cp$-even MSSM Higgs-boson mass $\Mh$
  has to be larger than its experimental limit $114.4 \gev$
  \cite{lephiggs,LEPHiggsSM}.%
  \footnote{%
  The limit on the SM Higgs mass
  holds unchanged for the light MSSM Higgs mass $\Mh$ for
  $\MA \gsim 150 \gev$. For lower values of $\MA$, the bound on $\Mh$ is
  smaller but the restrictions implied on the $\mu$ and $A$ parameters
  are significant also in this case.%
  }
  $\Mh$ has been evaluated
  with \fh 2.0~\cite{feynhiggs}, based on
\citeres{mhiggsletter,mhiggslong,mhiggsAEC}.

\item The $\Stop/\Sbot$-contribution to the $\rho$
  parameter, evaluated up to the \twol\ level~\cite{delrhosusy2loop},
  does not exceed its experimental bound.

\item The branching ratios $\br(B_s\to\mu^+\mu^-)$~\cite{bsmumu} and 
  $\br(B \to X_s\ga)$~\cite{bsg} 
are in agreement with their experimental limits.%
\footnote{
We are grateful
to A.\ Dedes and G.\ Hiller for providing the respective codes.
}%
\end{itemize}
In order to be able to check the sensitivity on these bounds we use a
stronger and a weaker version for each bound, see \refta{tab:bounds}.
The two Higgs mass bounds take into account a $3 \gev$ uncertainty
due to unknown higher-order corrections~\cite{mhiggsAEC}, the weak
bound in addition an uncertainty of $5 \gev$ due to the imperfect
knowledge of the top mass~\cite{tbexcl}, on 
which $\Mh$ is much more sensitive than $\amu$. The two bounds on
$\De\rho^{\SU}$, $\br(B\to X_s \ga)$, and $\br(B_s\to\mu^+\mu^-)$
correspond to $2\si$ and $3\si$ bounds and to $90\%$ and $95\%$~C.L.\
bounds, respectively. 

\begin{figure}[htb!]
\BC
\epsfig{figure=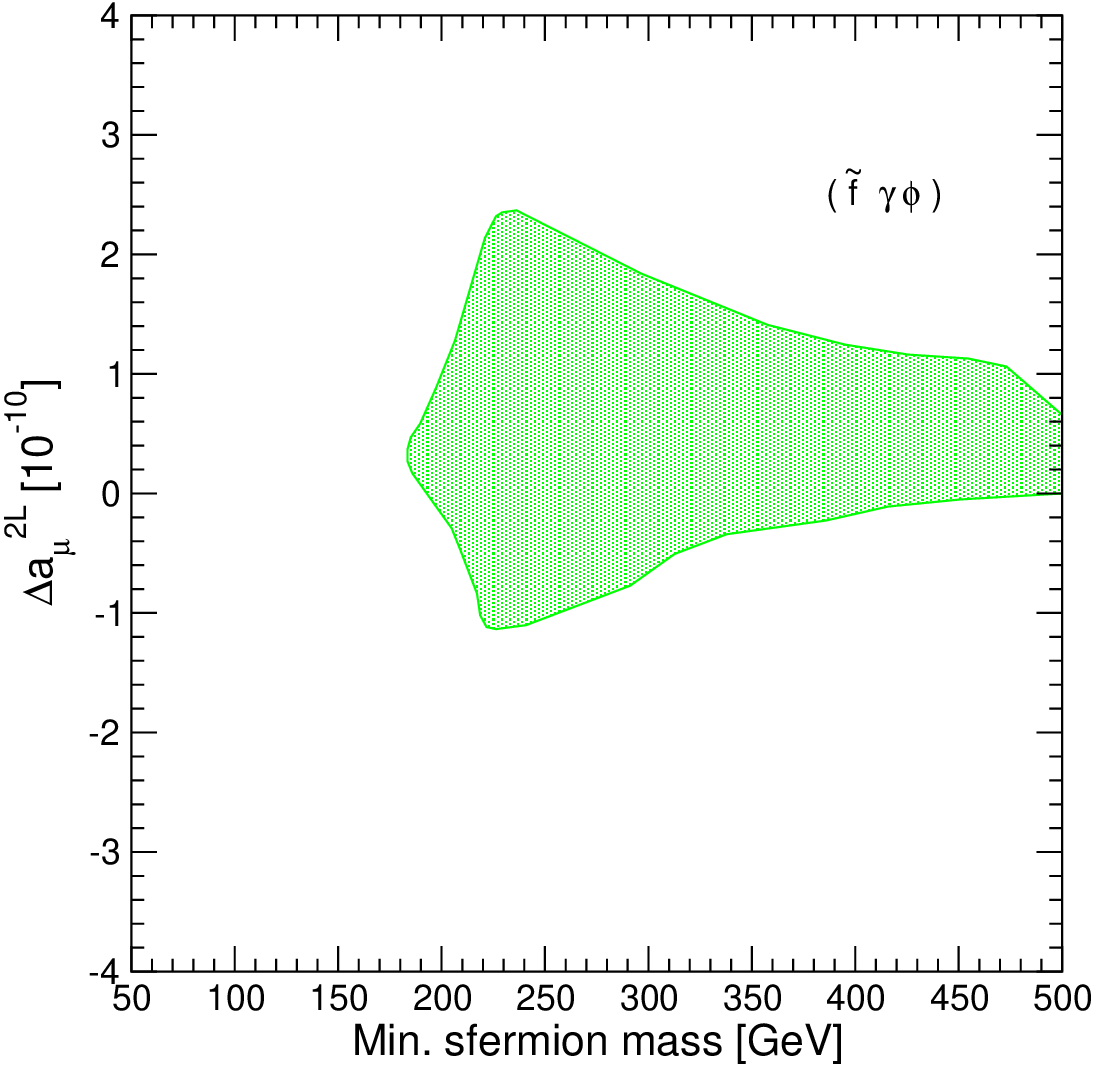,  width=7.5cm, height=5.5cm}~~~
\epsfig{figure=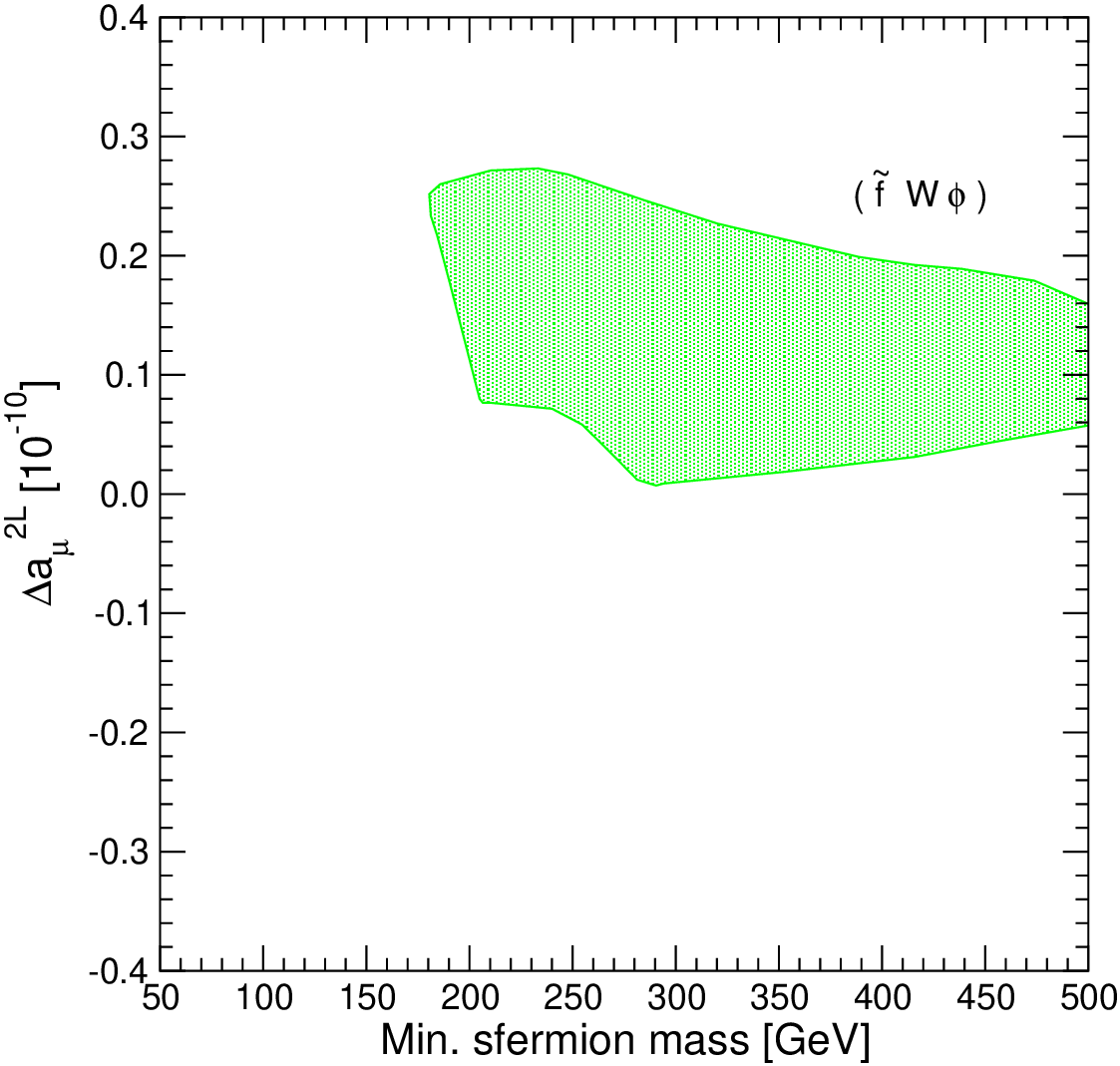,      width=7.5cm, height=5.5cm}\\[1.5em]
\epsfig{figure=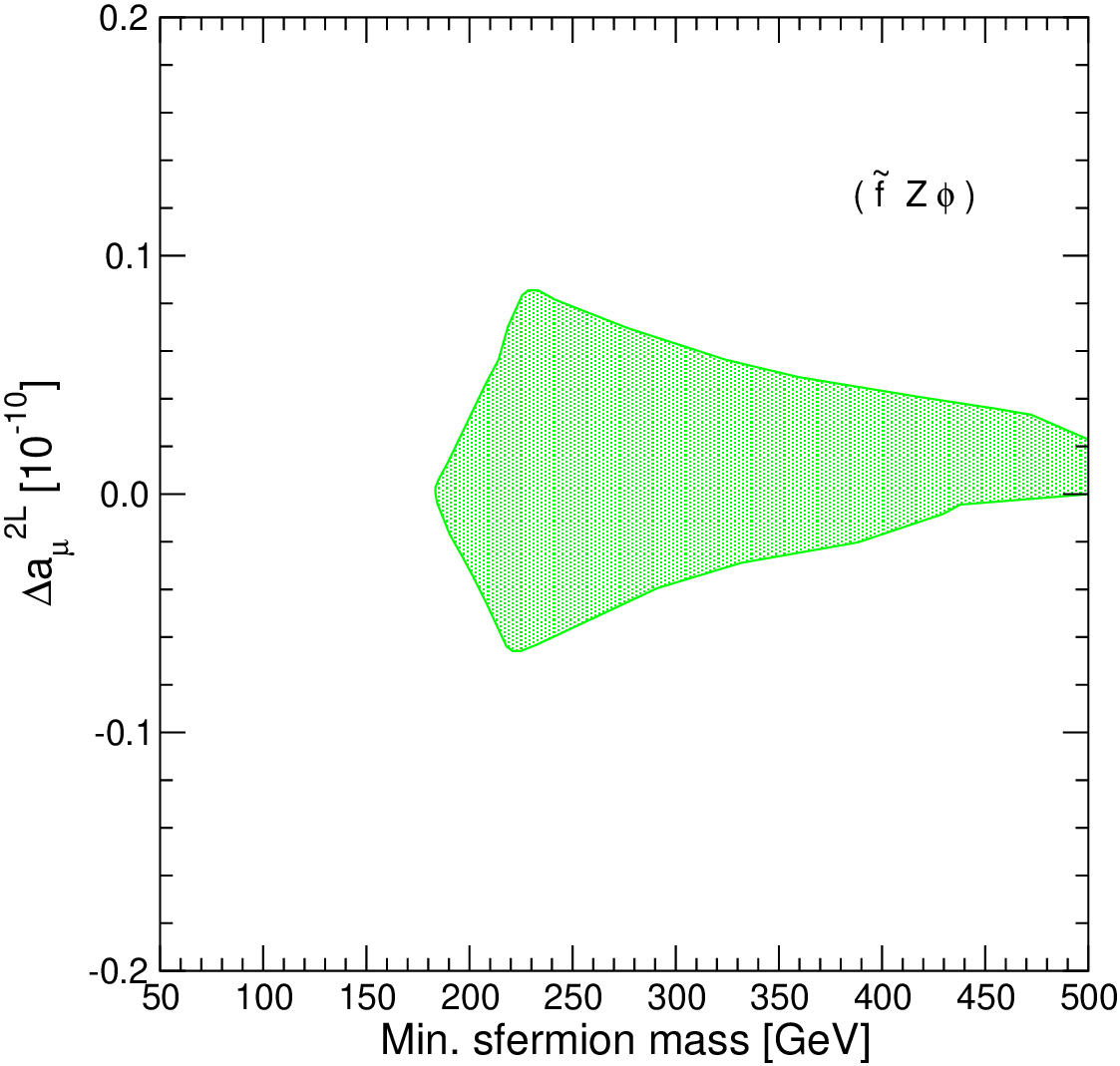,      width=7.5cm, height=5.5cm}~~~
\epsfig{figure=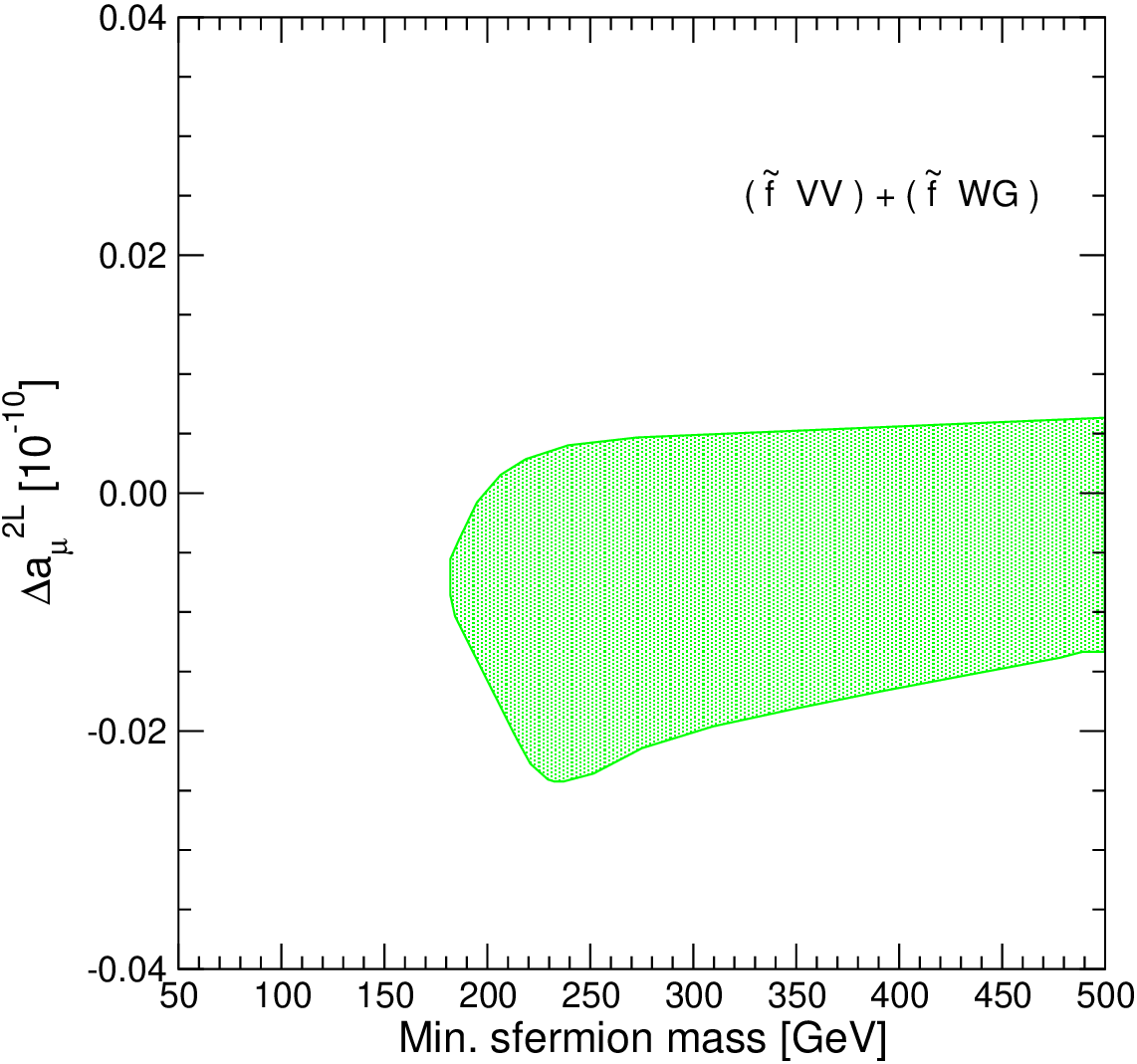, width=7.5cm, height=5.5cm}\\[1.5em]
\epsfig{figure=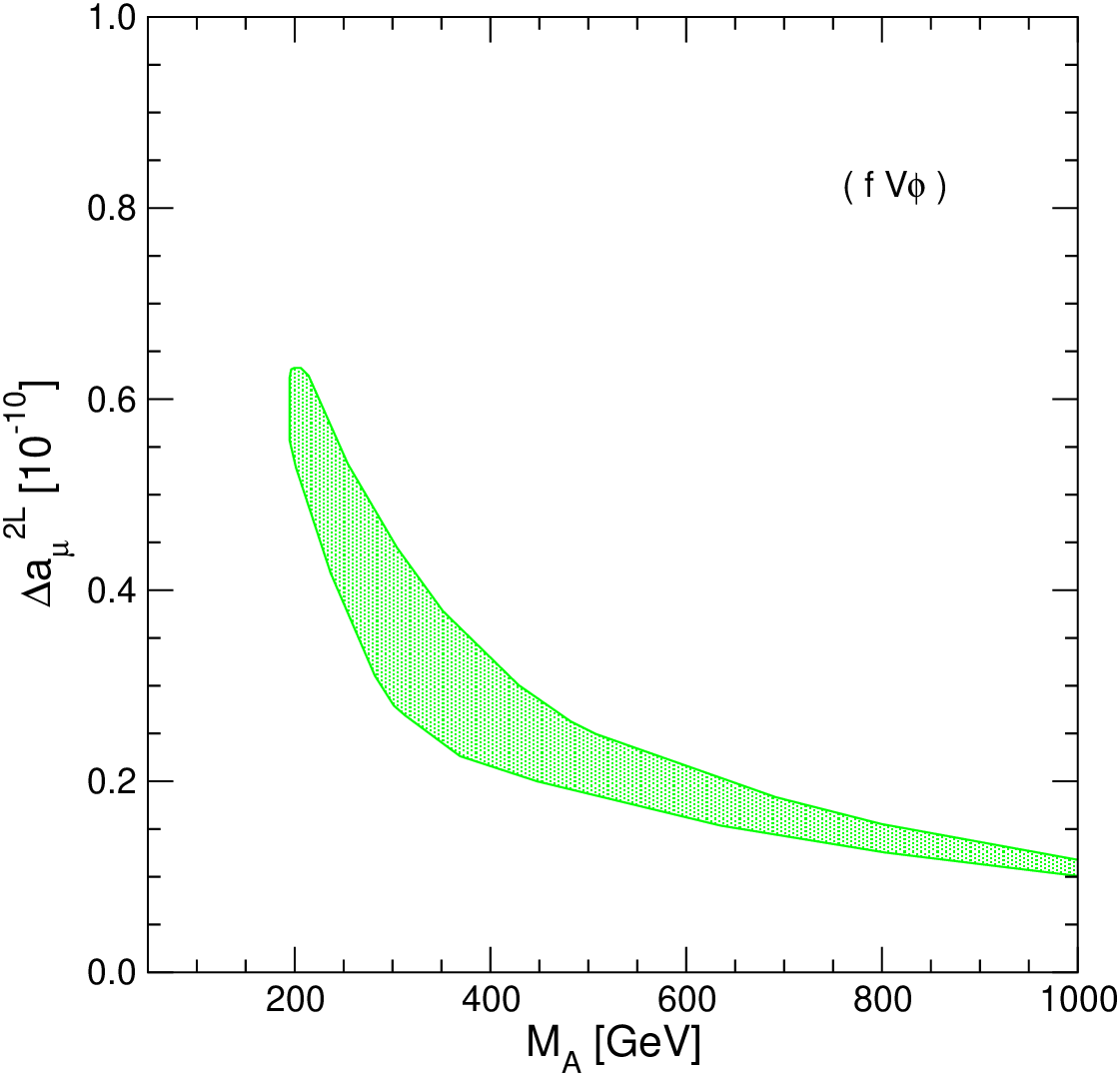,  width=7.5cm, height=5.5cm}
\\[0.5em]
~~~
\caption{%
Possible contributions to $\delamu$  for the case that 
all experimental bounds are required in their strong versions. The
results are subdivided into five classes of diagrams: sfermion loops
with gauge and Higgs boson exchange ($\sfn \ga \{h,H\}$), $(\sfn
W^\pm H^\mp$), ($\sfn Z \{h,H\}$), sfermion loops with gauge or
Goldstone boson exchange ($\sfn VV$)+($\sfn W^\pm G^\mp$), and fermion
loop diagrams ($f V \phi$). The results are plotted as functions of 
the lightest sfermion mass (sfermion loops) or $\MA$ (fermion loops).
}
\label{fig:scans}
\EC
\end{figure}

The interplay of these constraints restricts the allowed parameter
space severely. The $\Mh$ bound puts a limit of about 2.5$\msusy$ 
on $|A_t|$. The data on $b$ decays constrain $\mu$ and $A$
parameters in particular for small $\MA$. $\De\rho$ restricts the mass
splittings in the $\Stop,\Sbot$ sectors and thereby also the $\mu$ and
$A$ parameters, which appear in the off-diagonal elements of the
squark mass matrices. 

In \reffi{fig:scans} we plot the resulting ranges of possible
contributions of the individual classes of diagrams for the case that
all bounds are required in their strong versions. In the case of
the sfermion loop contributions we plot the results for $\amu$ over
the lightest sfermion mass (min$\{ \mste, \mstz, \msbe, \msbz \}$), 
and in the case of the fermion loop
contributions we plot the results over $\MA$.

{}From \reffi{fig:scans} the following conclusions can be drawn:
\begin{itemize}
\item The contribution of the gauge and charged Goldstone boson
  exchange diagrams,\\
  ($\sfn W^\pm G^\mp$) + ($\sfn VV$), is very
  small. Its maximum size is about
  $\frac{\alpha}{2\pi}\amu^{\rm EW,1L}\approx 0.02\times10^{-10}$.
\item The contribution of the ($\sfn Z \{h,H\}$) diagrams with $Z$ and
  Higgs exchange is at most of the order $0.1\times10^{-10}$ and
  thus negligible compared to the present experimental error. The
  reason for this suppression compared to the photon and $W$ exchange
  diagrams is the factor $(1-4\sw^2)$ in the coupling of the $Z$ to
  muons.
\item The contribution of the ($f V \phi$) diagrams with a fermion loop
  can reach $0.6\times10^{-10}$ for small
  $\MA \lsim 200 \gev$; the fermion loop diagrams are thus not
  completely negligible.
\item The photon exchange diagrams with a sfermion loop ($\sfn
  \ga \phi$) are dominant; 
  the results of the $W$ exchange diagrams ($\sfn
  W^\pm H^\mp$) are much smaller. The photon
  exchange diagrams are the only ones that can contribute more than
  $1\times10^{-10}$, the $W$ exchange diagrams contribute up to
  $0.3\times10^{-10}$. The reason for the suppression of the $W$
  diagrams is not only the high value of $\MW$ but also the fact that
  the $W$ couples to two different sfermions, to $\Stop-\Sbot$ or to
  $\Stau-\Snet$, of which at least one is usually relatively heavy.
\end{itemize}
Hence we find that the photon exchange contributions
calculated in \citeres{g-2BarrZee2} are indeed the dominant subclass of
diagrams with a closed (s)fermion loop.
We also find, however, that the
maximum contributions of more than $20\times10^{-10}$ quoted in
\citeres{g-2BarrZee1,g-2BarrZee2} for the photon and $W$ contributions
are reduced  to about $2.5\times10^{-10}$ and $0.3\times10^{-10}$
due to the experimental constraints on the MSSM parameter space.
Owing to the smallness of these contributions, the fermion loop
contributions can make up for a non-negligible part of the
two-loop corrections.


\section{Leading contributions and influence of the experimental
  constraints}
\label{sec:expconst}

Let us now focus on the photon and $W$ exchange contributions $(\sfn \ga
\{h,H\})$, $(\sfn W^\pm H^\mp)$ and study
the influence of the individual constraints on the ranges of possible
numerical values. We choose this set of contributions not only because
the photon contributions are dominant within our class of diagrams,
but also because these
contributions are obviously significantly restricted by the
experimental constraints. In contrast, the fermion loop contributions
$(f V\phi)$ can be non-negligible but depend mainly on $\MA$ and are
hardly constrained.

It is instructive to explicitly discuss the complete expression for
the photon diagrams (see e.g.\ \citere{g-2BarrZee2}):
\BEA
\De\amu^{(\sfn \ga \phi),\rm2L} &=& -
\frac{\alpha}{\pi}\ \frac{G_\mu m_\mu^2}{8\sqrt2\pi^2}
\ 
\frac{(N_c Q^2)_{\sfn}\la_{\mu \phi}\la_{\sfn \phi}}{M_\phi^2}
\ {\cal F}\left(\frac{m_{\sfn}^2}{M_\phi^2}\right),
\EEA
where $\sfn$ can be one of $\tilde{t}_{1,2}$, $\tilde{b}_{1,2}$,
$\tilde{\tau}_{1,2}$, and $\phi$ can be one of the $\cp$-even Higgs
bosons, $h$ or $H$. The couplings $\lambda$ are defined as ($\sa=\Sa$,
$\ca=\Ca$, etc.)
\BEA
\la_{\mu\{h,H\}} &=&  \{-\sa, \ca\}/\cbe\\
\la_{\Stop_i\{h,H\}} &=&  
 2\mt\Big(\mu\{\sa,-\ca\}+A_t\{\ca,\sa\}\Big)
 U^{\Stop}_{i 1}U^{\Stop}_{i 2}/\sbe
\non\\&& 
+\frac{6 \cw \mt^2\{\ca,\sa\} +  \MW \MZ\sbe(3 - 4\sw^2)\{-\sab,\cab\}}
      {3\cw \sbe}(U^{\Stop}_{i 1})^2
\non\\&&
+ 
\frac{ 6\cw\mt^2\{\ca,\sa\} + 4\{-\sab,\cab\}\MW\MZ\sbe \sw^2}
     {3\cw\sbe}(U^{\Stop}_{i 2})^2
\\
\la_{\Sbot_i\{h,H\}} &=& 
 2\mb\Big(-\mu\{\ca,\sa\}+\Ab\{-\sa,\ca\}\Big)
 U^{\Sbot}_{i 1}U^{\Sbot}_{i 2}/\cbe
\non\\&&
+\frac{6 \cw \mb^2\{-\sa,\ca\} +  \MW \MZ\cbe(-3 + 2\sw^2)\{-\sab,\cab\}}
      {3\cw \cbe}(U^{\Sbot}_{i 1})^2
\non\\&&
 + 
\frac{ 6\cw\mb^2\{-\sa,\ca\} - 2\{-\sab,\cab\}\MW\MZ\cbe \sw^2}
     {3\cw\cbe}(U^{\Sbot}_{i 2})^2
\EEA
and similar for $\lambda_{\tilde{\tau}\{h,H\}}$. The matrices
$U^{\Stop,\Sbot}$ diagonalize the sfermion mass matrices $M_{\sfn}^2$ in
the form $U^{\sfn} M_{\sfn}^2
(U^{\sfn})^\dagger=$ diag($m_{\sfn_1}^2,m_{\sfn_2}^2$). The loop function
${\cal F}$ is given by 
\BEA
{\cal F}(z) &=& \int_0^1 dx\frac{x(1-x)\log[z/(x(1-x))]}
{z-x(1-x)}.
\EEA
The result for the $W$ contribution has a similar form. 

This type of
contributions can be particularly enhanced by the ratio of the mass
scale of the dimensionful Higgs--Sfermion coupling
divided by the mass scale of the particles running in the loop, i.e.\
by ratios of the form
$\{\mu,A,\frac{\mt^2}{\MW}\}/\{m_{\sfn},M_{h,H}\}$, which can be much
larger than one. For
large $\tb$ and large sfermion mixing, the leading 
terms are typically given by the parts of the couplings with the
highest power of $\tb$ 
and by the 
loop with the lightest sfermion. 
These contributions involve only $H$-exchange, since the
$h$-couplings approach the SM-Higgs coupling for not too small $\MA$.
They can be very well  approximated by the formulas
\BEA
\label{stopcontrib}
\De\amu^{\Stop,{\rm 2L}} &=&
-0.013\times10^{-10}\;\frac{\mt\, \mu \tb}{\mst \MH}{\rm\ sign}(\At), \\
\De\amu^{\Sbot,{\rm 2L}} &=&
-0.0032\times10^{-10}\;\frac{\mb\, \Ab \tan^2\beta}{\msb \MH}{\rm\ sign}(\mu),
\label{sbotcontrib}
\EEA
where $\mst$ and $\msb$ are the masses of the lighter $\Stop$ and
$\Sbot$, respectively, and $\MH$ is the mass of the heavy $\cp$-even
Higgs boson. The formulas use the approximation ${\cal
  F}(m_{\sfn}^2/\MH^2)/\MH^2\approx0.34/(m_{\sfn}\MH)$ for the loop
function, which holds up to few percent if the respective
sfermion mass fulfils $m_{\Stop,\Sbot}\lsim\MH$. 
Since the heavier sfermions also contribute and
tend to cancel the contributions of the lighter sfermions, these
formulas do not approximate the full result very 
precisely, but they do provide the right sign and order of magnitude.

Equations (\ref{stopcontrib}), (\ref{sbotcontrib}) show that the
$\mt$-contributions are enhanced by one power of $\tb$ from the muon
Yukawa coupling and by the ratio $\mu/\MH$, whereas the
$\mb$-contributions contain an additional power of $\tb$ from the $b$
Yukawa coupling and the ratio $\Ab/\MH$. For $\tb=50$ and $\mu$ and
$A$ parameters larger than $1 \tev$ both contributions can amount to more
than $1\times10^{-10}$. However, $\mu$ is much more constrained by the
four experimental bounds in \refta{tab:bounds} than $\Ab$. Therefore,
the largest 
contributions in \reffi{fig:scans} originate from the sbottom loop
diagrams and from parameter constellations where ${\Sbot_1}$ is the
lightest sfermion.

\begin{figure}[htb!]
\BC
\epsfig{figure=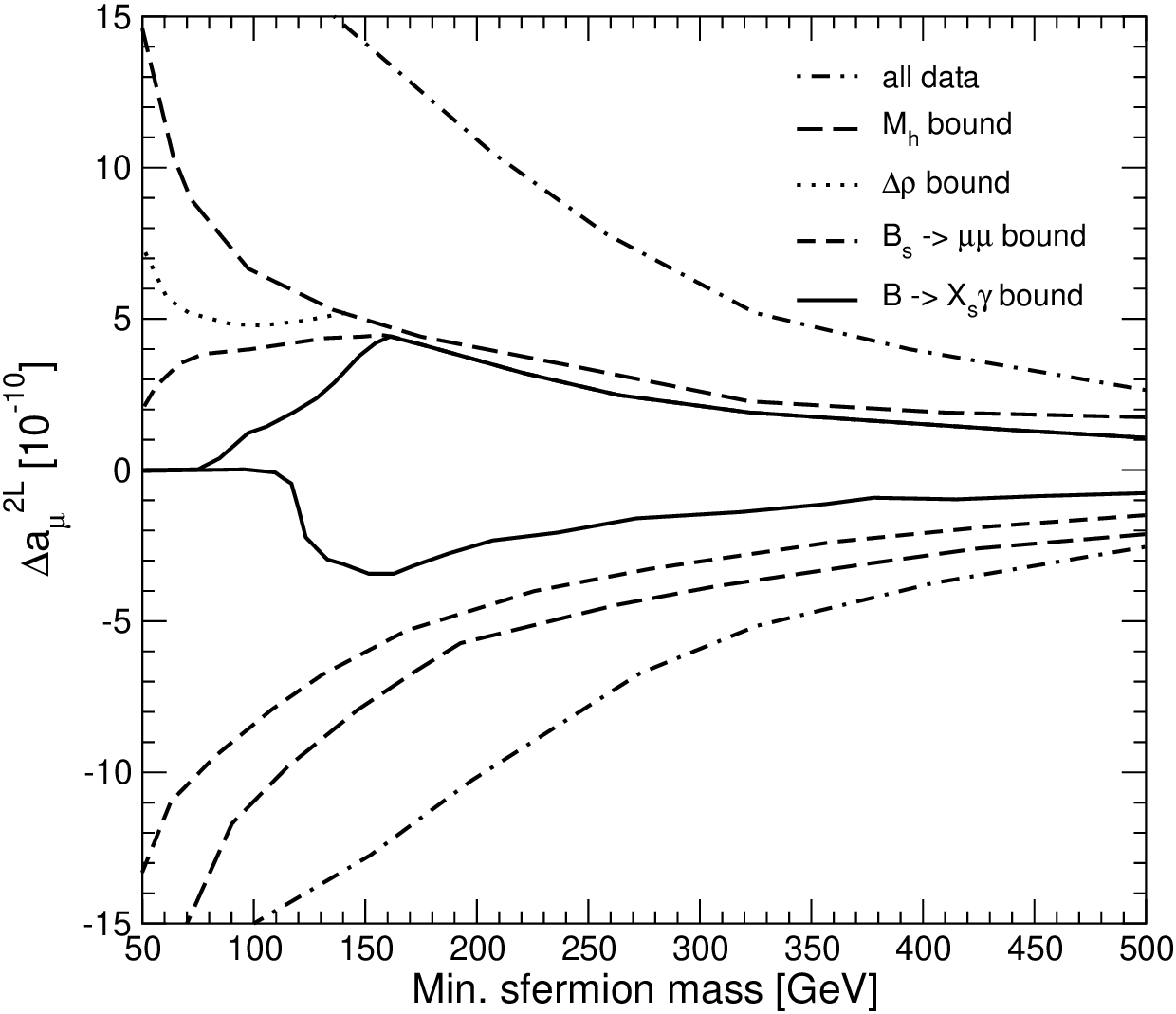, width=11cm}
\caption{%
Maximum contributions of the ($\sfn \ga \{h,H\}$) and ($\sfn W^\pm H^\mp$)
diagrams to $\delamu$ as a function of the lightest squark mass, 
min\{$\mste$, $\mstz$, $\msbe$, $\msbz$\}. No constraints except for the
parameter ranges in eq.\
(\ref{bounds}) are taken into account for the outermost curve. Going
to the inner curves additional weak constraints (see text) have
been applied.
}
\label{fig:gammaWscan}
\EC
\end{figure}

\begin{figure}[htb!]
\BC
\epsfig{figure=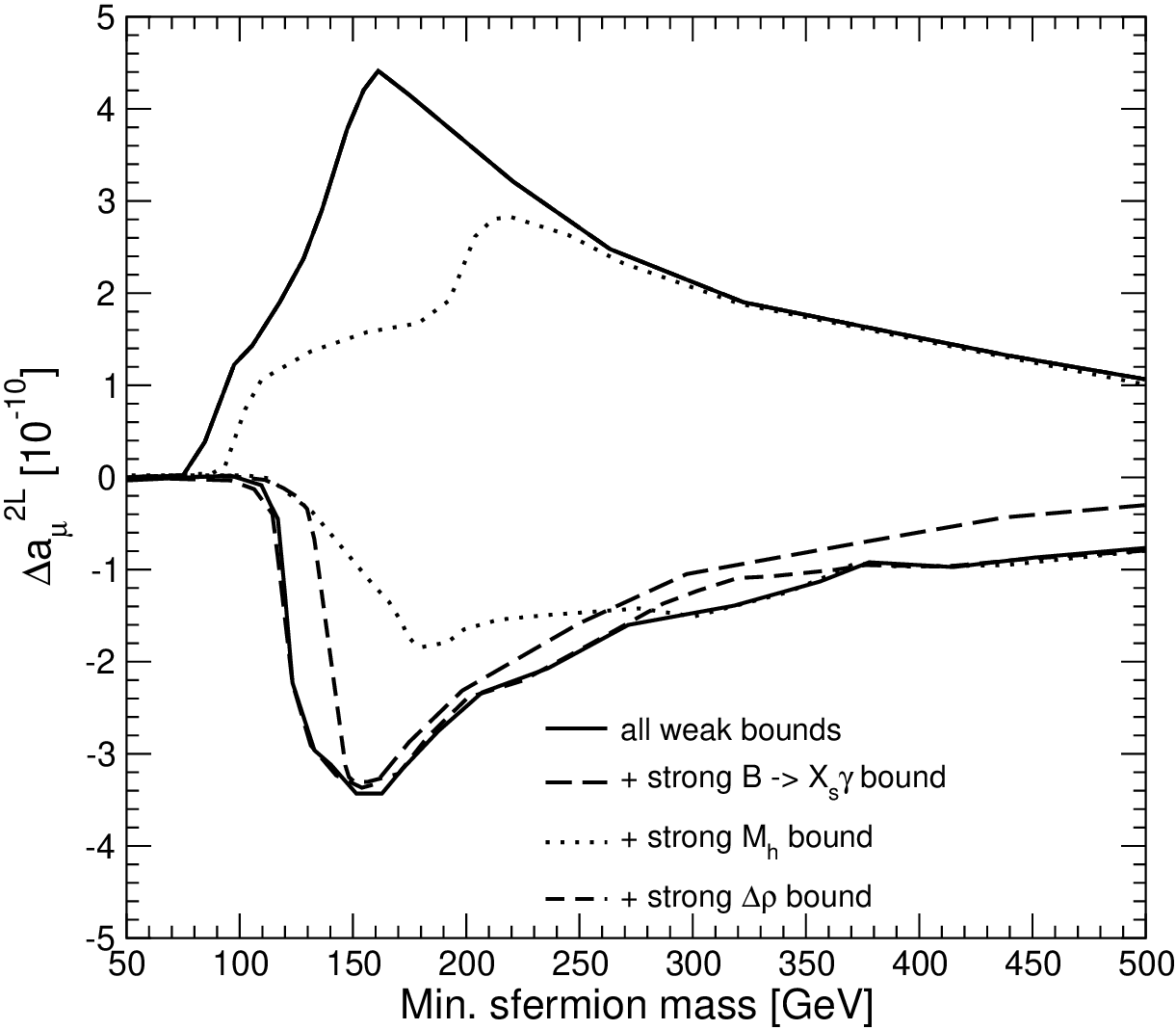, width=11cm}
\caption{%
Maximum contributions of the ($\sfn \ga \{h,H\}$) and ($\sfn W^\pm
H^\mp$) 
diagrams to $\delamu$ as a function of the lightest squark mass, 
min\{$\mste$, $\mstz$, $\msbe$, $\msbz$\}. The outer curve corresponds
to weak bounds for all experimental constraints. Each inner curve
takes into account one additional stronger constraint. Strengthening the
$B_s\to\mu^+\mu^-$-bound has a very small impact and 
is not shown. The inner area consequently corresponds to all strong
constraints.
}
\label{fig:gammaWscan_strong}
\EC
\end{figure}

Now we study the influence of the individual experimental
constraints on the photon and $W$ exchange contributions. 
Figure~\ref{fig:gammaWscan} is based on a data  
sample of $\sim300000$ parameter points in the range specified in
\refeq{bounds}, on which the weak versions of the bounds in
\refta{tab:bounds} are
incrementally applied. Figure \ref{fig:gammaWscan_strong} is based on
the data points satisfying all weak constraints and shows the effect
of strengthening each bound separately.

The results shown in Fig.\ \ref{fig:gammaWscan} are the
following:
\begin{itemize}

\item The outer lines show the largest possible results if all
  experimental bounds are ignored. They show a steep rise of $\delamu$
  for decreasing $m_{\sfn_1}$; for $m_{\sfn_1} < 150 \gev$ contributions
  larger than $15 \times 10^{-10}$, corresponding to two standard
  deviations of the experimental error on $\amu$, are possible.

\item The next two lines show the possible results if the bound
  $\Mh > 106.4\gev$ and then in addition the bound on $\De\rho$ are
  satisfied. The maximum contributions are very much reduced already
  by the $\Mh$ bound, and the $\De\rho$ bound reduces further the
  positive region for small sfermion masses. If both bounds are taken
  into account,   $\delamu > 5 \times 10^{-10}$ and  $\delamu < -10 \times
  10^{-10}$  is excluded for $m_{\sfn_1} \gsim 100 \gev$.

\item The two innermost lines correspond to taking into account in
  addition the bound on $\br(B_s\to\mu^+\mu^-)$ and finally also on 
  $\br(B \to X_s\ga)$. In particular taking into account the 
  $\br(B\to X_s\ga)$ bound 
  eliminates most data points with $m_{\sfn_1} \lsim 150 \gev$ and thus
  leads to a strong reduction of the possible size of the
  contributions. The largest 
  contributions of $\pm 4 \times10^{-10}$ to $\delamu$, corresponding to 
  $\sim 0.7\si$ of the experimental error, are possible for
  $m_{\sfn_1} \approx 150 \ldots 200 \gev$.

\end{itemize}
In applying all bounds one should be aware that any flavour
non-universality in the MSSM parameters could have a strong effect on
the predictions for the $b$ decays, whereas the influence on $\Mh$ and
$\De\rho^{\SU}$ would be mild. Hence it is interesting that even if
the $b$ physics bounds are ignored and only the weak $\Mh$ and
$\De\rho$ bounds are taken into account, the largest possible
contributions are strongly restricted to
\BE
-10\times10^{-10}<\delamu<5\times10^{-10}
\EE
for sfermions heavier than $100 \gev$.

Figure \ref{fig:gammaWscan_strong} shows that strengthening the bound
on $\Mh$ from $\Mh > 106.4 \gev$ to $\Mh > 111.4 \gev$ has the most
significant effect. It cuts off all the regions where
$\delamu > 3 \times 10^{-10}$ and $\delamu < -2 \times
10^{-10}$. Strengthening the other bounds has only a marginal
effect. This confirms that the 
$\Mh$-bound is most important for  restricting the parameter space.


\section{Non-universal soft SUSY-breaking parameters}
\label{sec:nonuniv}

Up to now we have found only moderate numerical effects from the two-loop 
diagrams with a closed (s)fermion loop, even for the photon
exchange diagrams. However, the approximation formula
eq.\ (\ref{stopcontrib}), $\De\amu^{\Stop,\rm 2L}\propto \mu
\mt/(m_{\Stop} \MH)$, 
shows that values up to $15\times10^{-10}$ should
be possible if $\mu\sim3\tev$ and $m_{\Stop}$, $\MH\sim150\gev$ (for
$\tb=50$). In Fig.\ \ref{fig:gammaWscan} such contributions indeed
appear but they are excluded if the experimental constraints are taken
into account. Already imposing only the bound on $\Mh$ reduces the
maximum contributions almost by a factor of 3.

It is important to keep in mind that all results presented so far were
based on the universality assumption $M_Q=M_U=M_D=M_L=M_E=\msusy$ for the
soft SUSY-breaking parameters in the MSSM.\footnote{%
SU(2) gauge invariance dictates only that the same left-handed squark
mass parameter $M_Q$  appears in the stop and sbottom mass matrices (and
analogously for $M_L$ in the slepton sector). Apart from this, there is
a priori no reason in the unconstrained MSSM to assume equality of the
left- and right-handed sfermion mass parameters, except for
simplicity. The symmetries of the MSSM allow independent values for
$M_{Q,U,D,L,E}$.
}
In this section we examine the
effect of relaxing this assumption. We do not attempt a full scan of
the MSSM parameter space but rather investigate which pattern of
non-universality 
can lead to particularly large results.

The reason why the large results in Fig.\ \ref{fig:gammaWscan} are
excluded is that universality indirectly leads to severe constraints
on the $\mu$ parameter. Via universality, the left- and right-handed 
diagonal elements of the 
stop and sbottom mass matrices are linked, and if one requires a light
stop, there is not much room for an even lighter sbottom. Hence the
off-diagonal element in the sbottom sector cannot be much larger than
the one in the stop sector, which means for large $\tb$, 
$\tb \gsim \mt/\mb$:
\BE
|\mu| \lsim |\At|.
\label{mueconstraint}
\EE
$\At$ is not only restricted by the requirement that all stop squared
masses are positive but also by the $\Mh$-bound, which roughly leads 
to $|\At|\lsim2.5\msusy$. Light stops require $|\mt\At|\approx\msusy^2$
and are therefore only possible for 
$\msusy \lsim 400 \gev$, and thus $|\At|$ can hardly exceed $1
\tev$. Because of universality and eq.\ (\ref{mueconstraint}), the
bounds on $\At$ hold also for $\mu$, and therefore also $\mu\lsim1\tev$.

It is therefore interesting  to break up  the relation between the stop
and sbottom mass parameters and to require only
\BE
\msusy=M_Q=M_U=M_L\ne M_D=M_E.
\EE
Thus we can choose small values of $\msusy$ and $\At$, giving rise to a 
light stop. Choosing $M_D\gg\msusy$ at the same time allows very large
$\mu$ without producing a too light sbottom. We will see that these
large values of $\mu$ are also compatible with the bound on $\Mh$.

\begin{figure}[t!]
\BC
\epsfig{figure=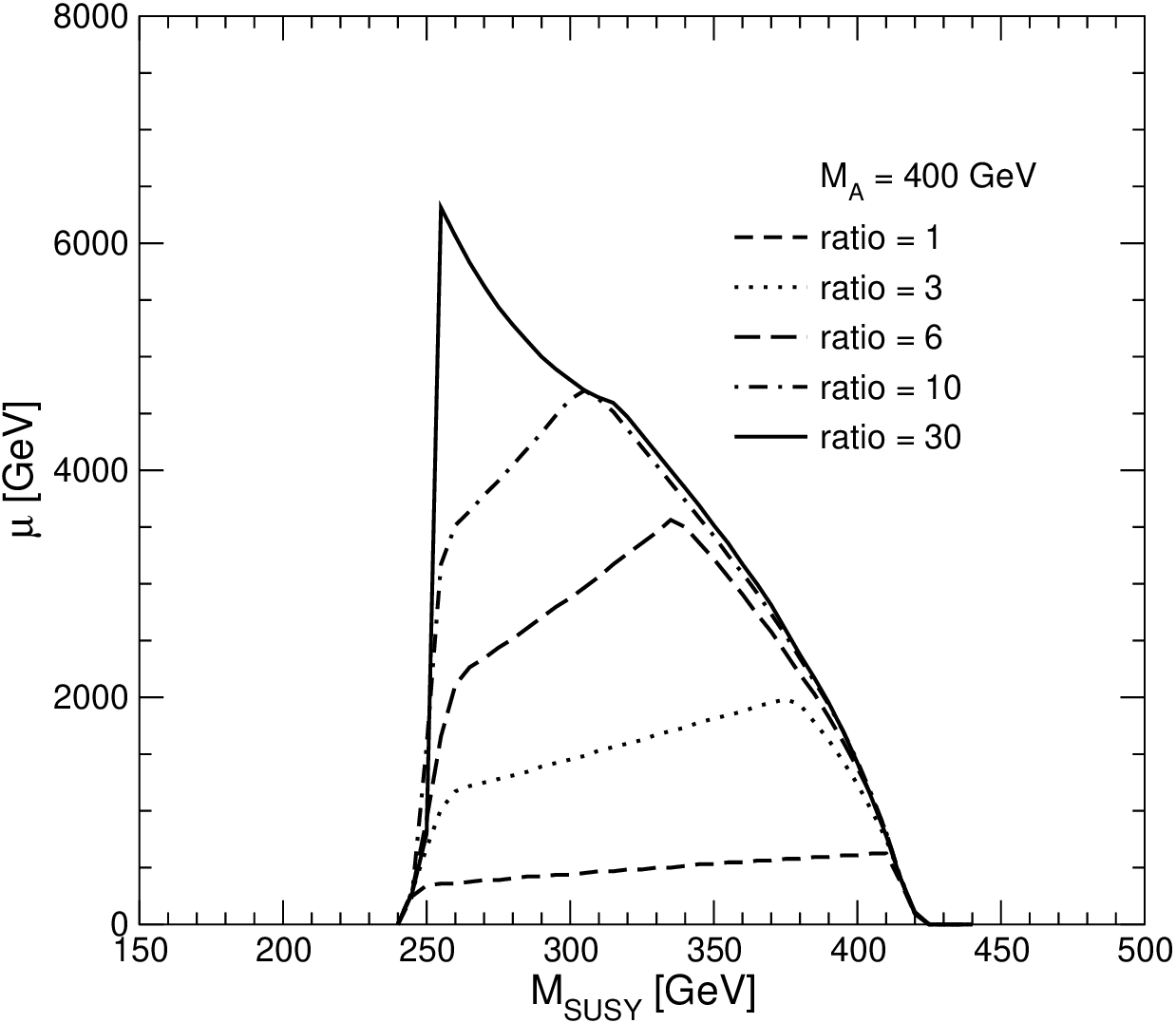, width=9.75cm,height=7.5cm}\\[1.5em]
\quad\hspace{-1.5mm}\epsfig{figure=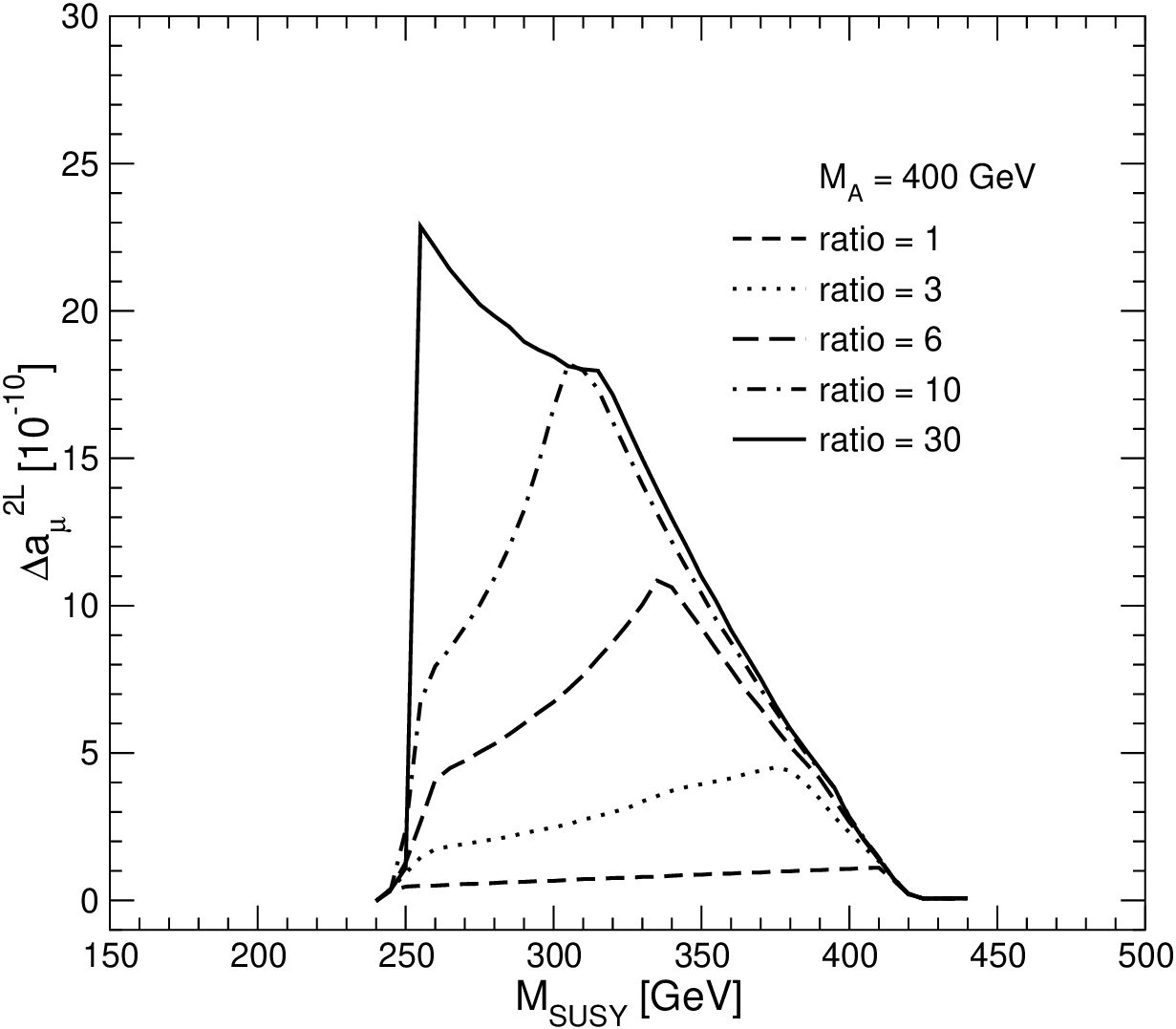, width=9.5cm,height=7.5cm}
\caption{%
Possible values of $\mu$ (upper plot) and corresponding contributions
to $\delamu$ (lower plot) for the case of
non-universality of the soft SUSY-breaking parameters.
The plots show $\mu$ and $\delamu$ as a
function of $\msusy=M_Q=M_U$ for different values of the ratio
$M_D/\msusy$. The other parameters are chosen as
$m_{\Stop_1}=150\gev$, $\MA=400\gev$, $\Ab=0$, $\tb=50$.
}
\label{fig:nonuniversality}
\EC
\end{figure}

\reffi{fig:nonuniversality} shows the results for different
ratios of $M_D/\msusy$. 
We choose a light stop mass $m_{\Stop_1}=150\gev$ and a moderate value
$\MA=400\gev$ in order to avoid too strong restrictions from $b$
decays. For each $\msusy$, $\At$ is determined by
$m_{\Stop_1}=150\gev$. The values of $\mu$ are determined as the
maximum values compatible with $\Mh>111.4\gev$ and
$\De\rho^{\SU}<0.004$.\footnote{%
$\Ab$ is set to zero here since the sbottom contributions cannot be
  expected to increase significantly beyond $\sim5\times10^{-10}$, see
eq.\ (\ref{sbotcontrib}).}

The upper plot in 
\reffi{fig:nonuniversality} shows these maximum values of $\mu$
as functions of $\msusy$. They significantly increase with
$M_D/\msusy$. Already for $M_D/\msusy=3$, values for $\mu$ larger than
$1.5\tev$ are 
possible. For $M_D/\msusy=6$, $\mu=3\tev$ is possible, and for
$M_D/\msusy=30$, even $\mu=6\tev$ is possible.

The lower plot in \reffi{fig:nonuniversality} shows the
corresponding results of the photon exchange diagrams 
$\De\amu^{(\sfn \ga\{h,H\}),\rm2L}$. 
We choose $\At<0$ so that the contribution to $\amu$ is
positive. The results exhibit a clear correlation with the values of
$\mu$, and they are quite precisely given by the approximation
(\ref{stopcontrib}).%
\footnote{%
Owing to the large values of $\mu$ the loop
  corrections to the heavy $\cp$-even Higgs mass $\MH$ can be large,
  and $\MH$, which enters in eq.\ (\ref{stopcontrib}), can be
  significantly lower than $\MA$.%
}%
~Thus the maximum results with $\mu < 3 \tev$ are about $10 \times
10^{-10}$, and the results for $\mu = 6 \tev$ are larger than $20
\times 10^{-10}$.

It should be noted that these parameter choices are rather extreme and involve
vastly different mass scales for the MSSM parameters. As an example, the
largest results are obtained for $\msusy\sim300\gev$, $M_D\sim9\tev$,
$\At\sim550\gev$,  
$\mu\sim 6\tev$. 
We have checked that all points plotted in
\reffi{fig:nonuniversality} 
with $\msusy=260\ldots390\gev$, where $\mu$ and
$\delamu$ are large, satisfy not only the bounds on $\Mh$
and $\De\rho$ but also those on $B\to X_s\ga$ and
$B_s\to\mu^+\mu^-$. Only if smaller values, $\MA<400\gev$, are chosen,
strong  violations of the $b$ decay bounds occur. For larger
$\MA$, on the other hand, the $b$ decay constraints are less
restrictive, and even larger values for $\mu$ and $\delamu$ than in 
\reffi{fig:nonuniversality} are possible.


\section{Conclusions}
\label{sec:conclusions}

We have obtained results for MSSM \twol\ corrections  to the anomalous 
magnetic moment of the muon. The corrections consist of diagrams where 
a SM fermion or sfermion loop is inserted into a \onel\ diagram with
gauge- and/or Higgs-boson exchange. 
We have investigated  the importance of the individual contributions
and the impact of existing experimental constraints on the maximum
numerical results.

It has been found that the by far
most important of the considered 
diagrams are the ones with a sfermion loop and photon
and neutral Higgs exchange ($\sfn \ga \{h,H\}$). They contribute up to
about  $2.5 \times 10^{-10}$ in the
parameter space allowed by all experimental constraints. This value
has to be compared with the current experimental error of
$6\times10^{-10}$. The diagrams with sfermion loop and $W^\pm$/$H^\mp$
exchange ($\sfn W^\pm H^\mp$) and the fermion loop diagrams ($f V\phi$)
contribute up to $0.3 \times 10^{-10}$ and
$0.6 \times 10^{-10}$, respectively, while the remaining diagrams are
negligible.

Our second result is that taking into account existing experimental
constraints is crucial. We have carefully analyzed the impact of the
constraints on the lightest Higgs-boson
mass, $\De\rho$, $\br(B_s\to\mu^+\mu^-)$ and $\br(B\to X_s\ga)$. 
If the 
experimental constraints were ignored and the $\mu$ and $A$ parameters
were varied up to $3\tev$, contributions of more than
$15\times10^{-10}$, corresponding to $2.5\si$ of the experimental error,
would be possible from the two-loop diagrams. Already if 
only the experimental bounds on $\Mh$ and $\De\rho$ are taken into
account, the accessible parameter space for $\mu$, $\At$, and $\Ab$ is
severely restricted, and one obtains
$-10 \times 10^{-10} < \delamu < 5 \times 10^{-10}$. Taking into account
all constraints leads to the relatively small result of 
$\delamu \lsim 3 \times10^{-10}$.
This two-loop correction of $\sim 0.5 \, \si$ therefore gives rise to 
only a moderate shift of the one-loop SUSY result (which can easily
account for the $\sim 3 \, \si$ deviation between the SM prediction and
the data).

The results quoted above have been obtained under the assumption of 
universal soft SUSY-breaking parameters $M_Q = M_U = M_D$. If one allows
large mass splittings between these parameters, the considered MSSM
\twol\ contributions can have a significantly larger numerical effect
while the existing constraints are still satisfied. We have
analyzed the example of 
$M_D>M_Q=M_U$, which can give rise to particularly
large contributions to $\amu$. One needs large ratios $M_D/M_Q>3$ and
at the same time a light stop and extremely large $\mu$ in order to
obtain contributions that are significantly higher than
$5\times10^{-10}$. Though in principle possible, such parameter
constellations look quite artificial, and they should be viewed as an
illustration of how difficult it is to produce larger contributions to
$\amu$. Models with universality at some high scale typically lead
to approximate low-energy universality $M_Q\approx
M_U\approx M_D$ and $M_Q \gg M_L\approx M_E$, which would restrict the
allowed range for $\mu$ even more than low-energy universality.

The contributions presented in this paper involve the potentially large
enhancement factors $\{\mu,A\}/\{m_{\tilde{f}},\MH\}$ and constitute an
important part of the \twol\ contributions in the MSSM. Our  full
result is
included as a Fortran subroutine in the code \fh\ (see: 
{\tt www.feynhiggs.de}). It can also be obtained as a {\em Mathematica}
formula from the authors upon request. 

In order to reduce the
theoretical uncertainty of the MSSM prediction for $\amu$ further, the
remaining \twol\ contributions should be analyzed as well. The
technical tools developed in this paper allow such a study, and the
results will be presented in a forthcoming publication.


\subsection*{Acknowledgements}
We thank  A.~Arhrib, A.\ Dedes, K.\ Desch, W.~Marciano, and D.\ Nomura,
for interesting discussions and A.\ Dedes and G.\ Hiller for providing
their codes. D.S. thanks
M.~Steinhauser and A.~Freitas for useful discussions and
checks of \twol\ asymptotic expansion and reduction algorithms.
This work has been supported by the European Community's Human
Potential Programme under contract HPRN-CT-2000-00149 Physics at
Colliders.



\end{document}